\pdfoutput=1

\documentclass[a4paper,11pt]{article}

\usepackage{jheppub,fancyhdr,epstopdf,color,amsmath,cases,slashed,hyperref,subfigure}
\usepackage{amsfonts}
\usepackage{amssymb}
\usepackage{graphicx,graphics,epsfig}
\usepackage{geometry}
\usepackage[english]{babel}
\usepackage{mathtools}
\usepackage[utf8]{inputenc}
\usepackage[T1]{fontenc}
 \usepackage{float}
\usepackage{hyperref}
\usepackage{caption}
\usepackage{caption}
\usepackage{subfigure}  
\usepackage{mathrsfs}  
\usepackage{bbold}
\usepackage{comment}
\usepackage{float}
\usepackage{stackrel}
\usepackage{multirow}
\usepackage{slashed}

\def\noi{\noindent}

\def\l{\lambda}

\def\ba{\begin{array}}
\def\ea{\end{array}}
\def\bea{\begin{eqnarray}}
\def\eea{\end{eqnarray}}

\newcommand{\Asla}{\not{\hbox{\kern-3.5pt $A$}}}
\newcommand{\Gsla}{\not{\hbox{\kern-3.5pt $G$}}}
\newcommand{\Wsla}{\not{\hbox{\kern-3.5pt $W$}}}
\newcommand{\Zsla}{\not{\hbox{\kern-3.5pt $Z$}}}

\newcommand{\Dslash}{\not{\hbox{\kern-4pt $D$}}}
\newcommand{\pslash}{\not{\hbox{\kern-2.3pt $p$}}}

\def\lsim{\;\raise0.3ex\hbox{$<$\kern-0.75em\raise-1.1ex\hbox{$\sim$}}\;}
\def\gsim{\;\raise0.3ex\hbox{$>$\kern-0.75em\raise-1.1ex\hbox{$\sim$}}\;}

\def\l{\lambda}

\def\ba{\begin{array}}
\def\ea{\end{array}}
\def\bea{\begin{eqnarray}}
\def\eea{\end{eqnarray}}

\def\lsim{\;\raise0.3ex\hbox{$<$\kern-0.75em\raise-1.1ex\hbox{$\sim$}}\;}
\def\gsim{\;\raise0.3ex\hbox{$>$\kern-0.75em\raise-1.1ex\hbox{$\sim$}}\;}

\newcommand{\beqn}{\begin{eqnarray}}
\newcommand{\eeqn}{\end{eqnarray}}

\title{Relic density of dark matter in the inert doublet model beyond leading order for the low mass region: 2. Co-annihilation }
\preprint{LAPTH-002/21, CERN-TH-2021-002}

\author[a]{Shankha Banerjee}
\author[b]{\!\!, Fawzi Boudjema}
\author[c, d]{\!\!, Nabarun Chakrabarty}
\author[e]{\!\!,  Hao Sun}

\affiliation[a]{CERN, Theoretical Physics Department, CH-1211 Geneva 23, Switzerland}
\affiliation[b]{LAPTh, Universit\'e Savoie Mont Blanc, CNRS, BP~110, F-74941 Annecy-le-Vieux, France}
\affiliation[c]{Centre for High Energy Physics, Indian Institute of Science, C.V. Raman Avenue, Bangalore 560012, India}
\affiliation[d]{Department of Physics, Indian Institute of Technology Kanpur, Kanpur, Uttar Pradesh 208016, India}
\affiliation[e]{Institute of Theoretical Physics, School of Physics, Dalian University of Technology, Dalian 116024, People’s Republic of China}

\emailAdd{shankha.banerjee@cern.ch}
\emailAdd{boudjema@lapth.cnrs.fr}
\emailAdd{chakrabartynabarun@gmail.com}
\emailAdd{haosun@dlut.edu.cn}

\abstract{We examine the relic density of the light mass dark matter region in the inert doublet model (IDM) when the dominant process is due to co-annihilation between the lightest neutral scalars of the model. The full one-loop electroweak corrections are computed in an on-shell scheme and are found to be well approximated as an effective cross-section expressed in terms of $Z$-observables. The electroweak corrections to the subdominant process which consists of an annihilation into an on-shell $W$ and an off-shell $W$, that is calculated as a annihilation into a 3-body final state, is also performed. The latter reveals an important dependence on a parameter that describes the self-interaction of the new scalars (solely within the dark sector), a parameter which is not accessible in tree-level calculations of standard model (SM)-IDM interactions.}

\begin{document}

\date\today

\maketitle


\section{Introduction}
\label{sec:coann_intro}
In the very thorough analysis we have conducted in Ref.~\cite{OurPaper1_2020} on the available parameter space of the inert ouble model (IDM)~\cite{Deshpande:1977rw, Barbieri:2006dq,Hambye:2007vf, LopezHonorez:2006gr,Cao:2007rm,Gustafsson:2007pc,Agrawal:2008xz,Hambye:2009pw, Lundstrom:2008ai, Andreas:2009hj, Arina:2009um, Dolle:2009ft, Nezri:2009jd, Miao:2010rg, Gong:2012ri, Gustafsson:2012aj, Swiezewska:2012eh, Arhrib:2012ia,Wang:2012zv, Goudelis:2013uca, Arhrib:2013ela, Krawczyk:2013jta, Osland:2013sla, Abe:2015rja,Blinov:2015qva, Diaz:2015pyv, Ilnicka:2015jba, Belanger:2015kga, Carmona:2015haa, Kanemura:2016sos,Queiroz:2015utg,Belyaev:2016lok,Arcadi:2019lka, Eiteneuer:2017hoh, Ilnicka:2018def, Kalinowski:2018ylg,Ferreira:2009jb,Ferreira:2015pfi,Kanemura:2002vm, Senaha:2018xek, Braathen:2019pxr, Arhrib:2015hoa,Garcia-Cely:2015khw,Banerjee:2016vrp,Basu:2020qoe,Abouabid:2020eik,Kalinowski:2020rmb}, for the low mass Dark Matter (DM)  (below the $W$ mass, $M_W$), we have confirmed the survival of a very small region with a DM mass of $55-60$ GeV which has recently been unraveled~\cite{Datta:2016nfz, Belyaev:2016lok}. A good value of the relic density~\cite{Ade:2015xua} is obtained. Within the freeze-out mechanism this is possible, mainly, through efficient co-annihilation. The two additional (lightest) neutral scalars of the IDM, $X$ and $A$, need to be thermodynamically close to one another for this co-annihilation to take place. In other words, this requires a certain degree of mass degeneracy which, in the context of the IDM, is not necessarily fine-tuned and unnatural. The $Z$-mediated co-annihilation is into the (light) fermion pairs, not including the top quark, of the Standard Model (SM). The process is a gauge interaction induced process and as such it is quite efficient. To obtain a value of the relic density around $0.12$, the co-annihilation cross-sections must be {\it dampened} by the Boltzmann factor, $e^{-m_-/T}$, where $m_-=M_A-M_X$ is the mass splitting and $T$ is the temperature. The Boltzmann factor in the  thermally averaged cross-sections should furnish enough reduction to counterbalance the large cross-sections. This in turn requires the mass splitting to be not too small. In our analyses we found that $m_-=8$ GeV is an optimal value. This value is at the edge of the LEPII constraint on the IDM. Higher values of the mass splitting would lead to visible tracks at LEP that disqualify the model while lower values do not give enough Boltzmann reduction leading to too small $\Omega h^2$. For our one-loop analyses, we have retained two benchmarks points with mass $M_X=58, 60$ GeV and $m_-=8$ GeV, see Table~\ref{tab:bpfbcoann}. 

\begin{table}[hbtp]
\centering
\begin{tabular}{|c|c |c |}
\cline{2-3}
\multicolumn{1}{c|}{}& P58&P60 \\
\hline
$M_X$  &   58&60\\ 
\hline 
$\l_L $  & 0.0&0.0 \\ \hline
$M_A$,$M_{H^\pm}$&      66, 110&68,150 \\  \hline 
$( \l_3,\l_4,\l_5)$ &  (0.28,-0.26,-0.02) & (0.60,-0.58,-0.02) \\ \hline
\hline
   \multicolumn{3}{|c|}{$\Omega h^2$} \\    \hline
$\alpha(0)$ &  0.113&0.116 \\    \hline
    $\Omega_{WW^\star}(\%)$      &5&9  \\ \hline
     $\Omega_{AX\to \sum f\bar f}(\%)$ & 95&91 \\ 
 \hline
 \end{tabular}
\caption{\label{tab:bpfbcoann}{\it Characteristics of the co-annihilation benchmarks points P58 and P60. All masses are in GeV. We also show tree-level (calculated with $(\alpha(0) $) relic density and the weight in percent of each channel contributing to the relic density. $f$ stands for all light fermions in the SM (the top-quark channel is closed). For more precise numbers, especially of the underlying parameters $\l_{3,4,5}$, please refer to~\cite{OurPaper1_2020}.}}
\end{table}

As we will see, this 2 GeV difference in the DM mass between the two benchmarks is enough to give a $20\%$ reduction in the (co)-annihilating cross-sections in P60 compared to P58. While the Boltzmann factor is (slightly) larger for the $M_X=60$ GeV case to somehow compensate for the drop in the $AX \to f \bar f$ cross-sections, what really benefits P60 is a larger cross-section of the annihilating process $XX \to WW^\star$, thanks to a larger threshold than in the case of $M_X=58$ GeV. The masses of all the scalars of the IDM have been set to define the model, an extra parameter that is needed to fully describe the interaction between these scalar and the SM scalar sector is $\l_L$. In fact, $\l_L$ describes the interaction strength of the SM Higgs to the DM, $X$, see~\cite{OurPaper1_2020} for details. 
We find that in order to keep the cross-sections small, including $XX$ annihilations, $\l_L$ must be as small as possible. This is the reason that for the surviving points we take $\l_L=0$. The latter suppresses SM Higgs exchange in $ XX \to W W^\star$ and avoids annihilation into $b \bar b$ through Higgs exchange. We must note that for sufficient relative velocity the SM Higgs resonance is accessed for this range of DM masses and even if $\l_L=0$ SM Higgs exchange may kick in at one-loop. The Higgs resonance scenarios deserve a dedicated analysis. Of course, the selection of the benchmark points was based on tree-level cross-sections. It is interesting to know how these cross-sections are affected by loop calculations. $A X \to f \bar f$ accounts for more than $95\%$ ($90\%$) of the contribution for the relic density for P58 (P60). In this section we therefore concentrate on the electroweak radiative corrections for such processes. The rest of the annihilation contributions are from $XX \to W W^\star$. In the present paper we include the results of the one-loop contribution to $XX \to W W^\star \equiv W f \bar f^\prime$ but we leave many of the technical details covering both $2 \to 3$ processes at one-loop and the treatment of the SM Higgs resonance to more detailed presentations in Refs.~\cite{OurPaper3_2020, OurPaper4_2020}.


\section{The co-annihilation cross-section $AX \to f \bar f$}
\subsection{$AX \to f \bar f$: tree-level considerations}
\begin{figure}[hbtp]
\begin{center}
\includegraphics[scale=0.3]{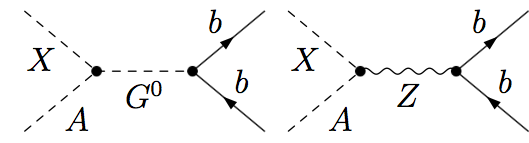}
\caption{\label{fig:coannff-treeaxbb}\it Tree-level Feynman diagrams (in the non-linear Feynman gauge\cite{Boudjema:1995cb,OurPaper1_2020}) for $AX \to b \bar b$ as a representative of $AX \to f \bar f$.}
\end{center}
\end{figure}
\noi The co-annihilation cross-sections are completely determined by the gauge coupling, given the masses of $X$ and $A$. The annihilations are essentially $P$-wave with a very small $S$-wave contribution that is noticeable only for extremely small velocities in the case of the heaviest final state fermion, the $b$-quark. For such $2\to 2$ processes, the cross-section can be described by a transparent analytical formula. With $m_f$ the mass of the final fermion, the tree-level cross-section can be written as 
 
\beqn
\sigma_{A X \to f \bar{f}} \; v =\sigma_{v^2}^{f\bar f} v^2+ \frac{m_f^2}{s}\sigma_0^{f\bar f}.
\eeqn

\begin{figure}[hbtp]
\begin{center}
\includegraphics[width=0.48\textwidth,height=0.36\textwidth]{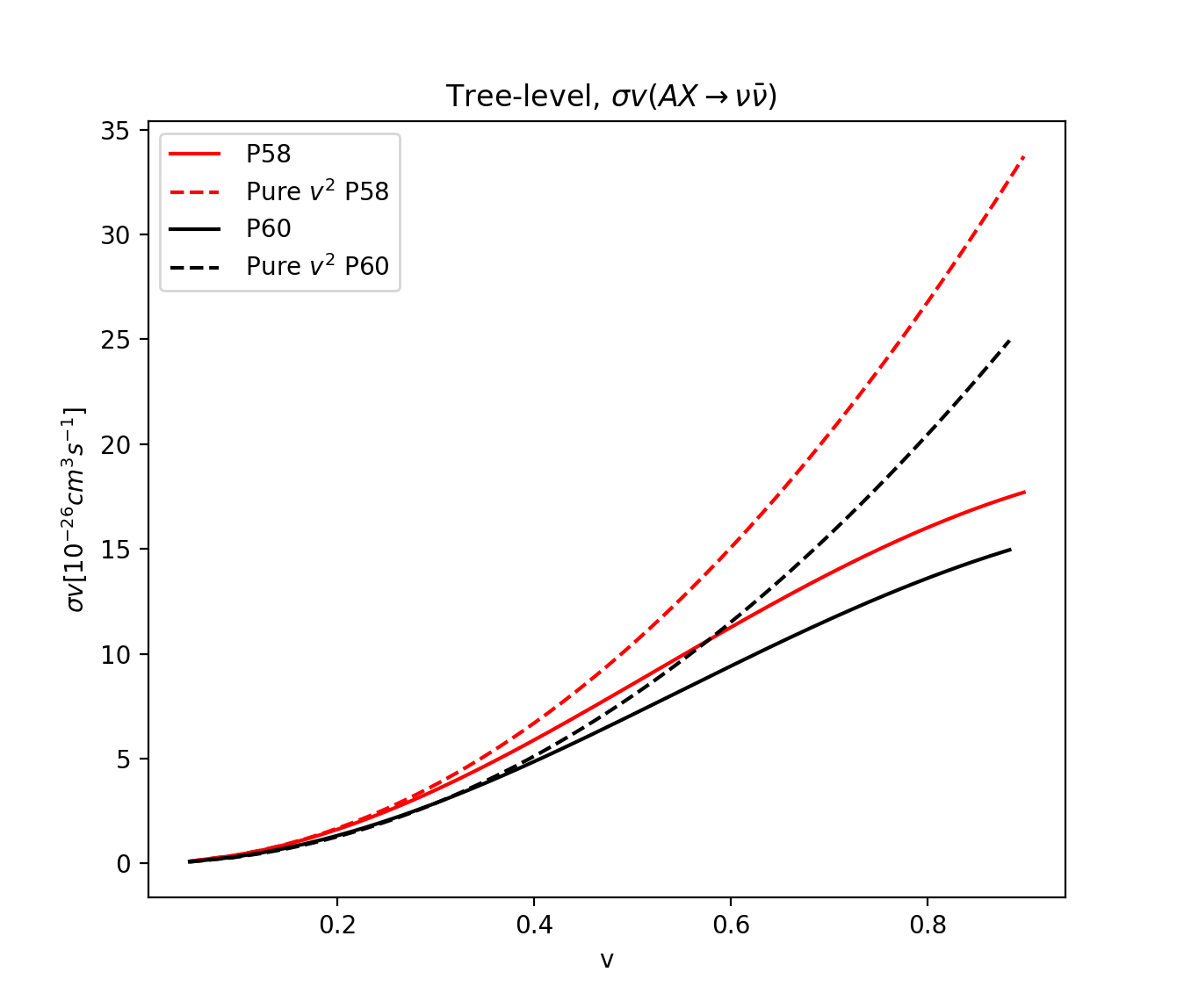}
\includegraphics[width=0.48\textwidth,height=0.36\textwidth]{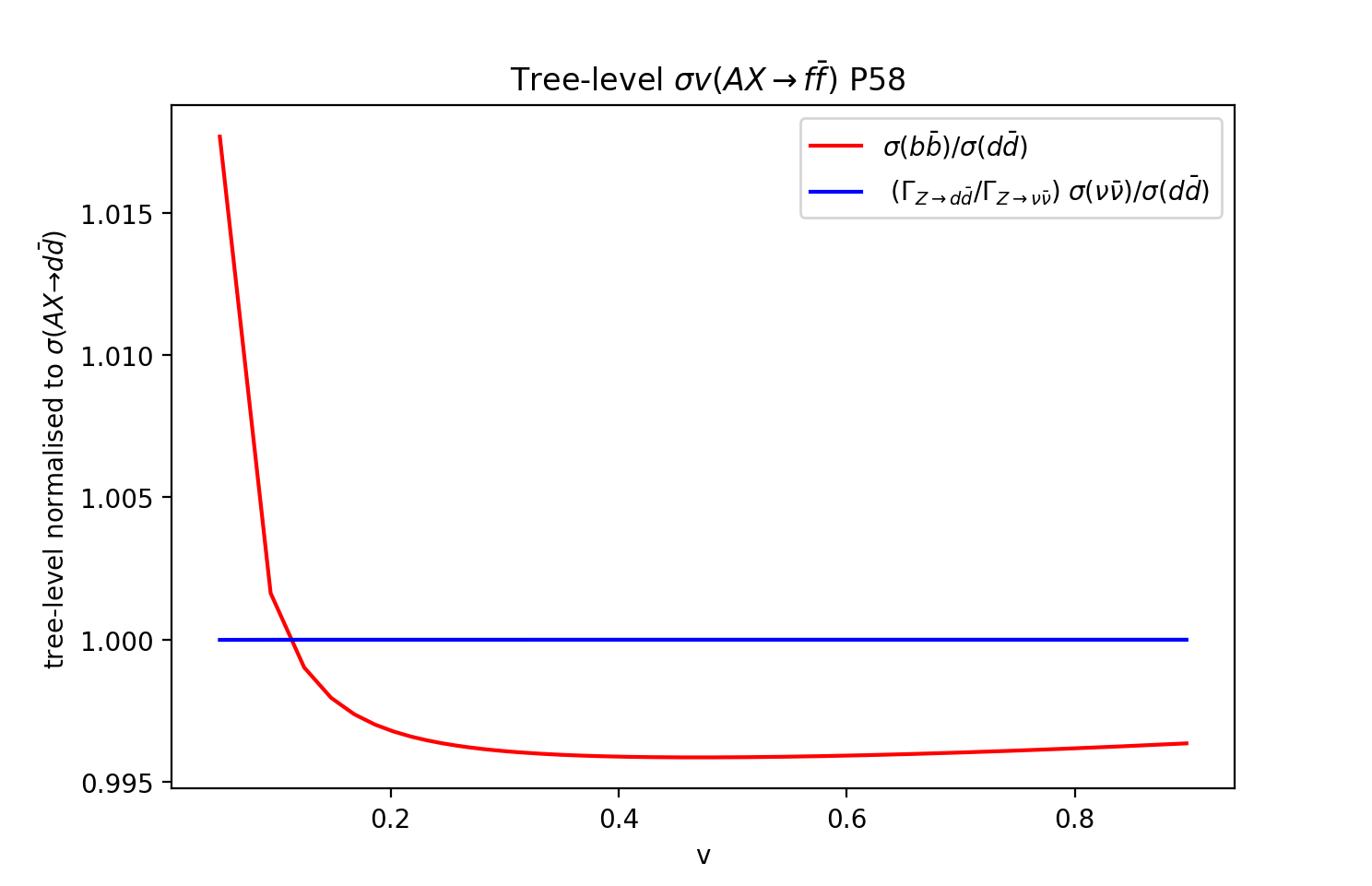}
\caption{\label{fig:coannff-xsff}\it {\underline{Left panel}}: The tree-level  cross-section for $\sigma (XX \to \nu \bar \nu)$ for P58 and P60 as a function of the relative velocity. The dashed curves are the pure $P$-wave ($\propto v^2$) approximation. We see that this pure $v^2$ approximation is only valid for $v < 0.4$. Note that the cross-sections are quite large, however, remember that a Boltzmann suppression will be applied to these co-annihilations cross-sections when they are converted to effective velocity/temperature averaged cross-sections.\\
\noi  {\underline{Right panel}}: The $b\bar b$ cross-section normalised to the $d \bar d$ cross-section and the $d \bar d$ cross-section normalised to the $\nu \bar \nu$ cross-section. See text for more detailed comments on these plots.}
\end{center}
\end{figure}

The $P$-wave, $v^2$ term, dominates by far while the very small $S$-wave contribution, $\sigma_0$, is suppressed by the chiral factor $m_f^2$. The latter is further suppressed by the mass splitting factor, $m_-^2$, as shown explicitly below. With  the ``{\it running}" partial width  
\beqn
\label{eq:zrunningwidth}
\Gamma_{Z \to f \bar f}(s)&=&N_c^f \frac{g^2 \sqrt{s}}{12 \pi c_W^2}  \beta_f \Bigg( g_V^2 \frac{3-\beta_f^2}{2}+g_A^2 \beta_f^2 \Bigg) \stackrel[ s \gg m_f^2]{ }
{\simeq} \frac{\sqrt{s}}{M_Z} \Gamma_{Z \to f \bar f}, \quad N_c \; \text{is the colour factor} \nonumber \\
&\text{and}& \quad \beta_f=\sqrt{1 - \frac{4 m_f^2}{s}}, \quad g_V = \frac{T_{3f}}{2} - Q_f s^2_W, 
\quad g_A = \frac{T_{3f}}{2} \quad (T_{3f}=\pm 1/2),
\eeqn
($Q_f$ is the charge of the fermion), we have 
\beqn
\label{eq:analytics-coannAX}
\sigma_{v^2}^{f\bar f}&=&\frac{g^2}{ c_W^2} \frac{1}{8 s}  \frac{1-m_-^2 m_+^2/s^2}{(1-M_Z^2/s)^2 
}   \frac{\Gamma_{Z \to f \bar f}(s)}{\sqrt{s}} \simeq
\frac{g^2}{ c_W^2} \frac{1}{8 s}  \frac{1-m_-^2 m_+^2/s^2}{(1-M_Z^2/s)^2 
}  \frac{\Gamma_{Z \to f \bar f}}{M_Z}\nonumber \\
\sigma_0^{f \bar f}&=&\frac{m_-^2 m_+^2}{s^2-m_-^2 m_+^2}\frac{g^2}{ c_W^2}  \frac{3 }{2 s} \frac{1}{(1-M_Z^2/s)^2}   \frac{\beta_f  \Gamma_{Z \to \nu \bar \nu}}{M_Z} \simeq \frac{m_-^2 m_+^2}{s^2-m_-^2 m_+^2}\frac{  g^2}{ c_W^2}  \frac{3 }{2 s} \frac{1}{(1-M_Z^2/s)^2}   \frac{ \Gamma_{Z \to \nu \bar \nu}}{M_Z}. \nonumber \\ 
\eeqn
The approximation ($\simeq$) in equation~\ref{eq:analytics-coannAX} which  amounts to $m_f \to 0$, is excellent for all fermions including the bottom-quark (for which the approximation is off by about 3 per-mille for the range of velocities relevant for the calculation of the relic density contributed by these channels). Despite our $P$-wave appellation, note that the $s$ (and hence $v$) dependence contained in $\sigma_{v^2}^{f\bar f}$ is not small for the 2 scenarios that we are studying, see Figure~\ref{fig:coannff-xsff}. This $s$-dependent factor in the expression of $\sigma_{v^2}^{f\bar f}$ explains that for the same value of the relative velocity, the co-annihilation processes are about 20\% larger for P58 as compared to P60. 
Figure~\ref{fig:coannff-xsff}, which is the result of a fully automated computation with our code {\tt SloopS}~\cite{Boudjema_2005, Baro:2007em, Baro:2008bg, Baro:2009na, Boudjema:2011ig, Boudjema:2014gza, Belanger:2016tqb, Belanger:2017rgu, Banerjee:2019luv}, exhibits the $P$-wave nature of the co-annihilation and confirms the analytical results. Departure from a pure $v^2$ dependence is clearly seen. The ratio $\sigma(b\bar b)/\sigma(d \bar d)$ clearly shows the very small contribution of the $S$-wave for small $v$ and the tiny effect of the mass of the fermion for larger $v$. In the case of massless fermions, Figure~\ref{fig:coannff-xsff} confirms that $\sigma(\nu \bar \nu)/\sigma(d \bar d)$ is given by the ratio of the respective partial widths of the $Z$ into these fermions.

\subsection{$AX \to f \bar f$: cross-sections at one-loop}
\begin{figure}[h]
\begin{center}
\includegraphics[scale=0.5]{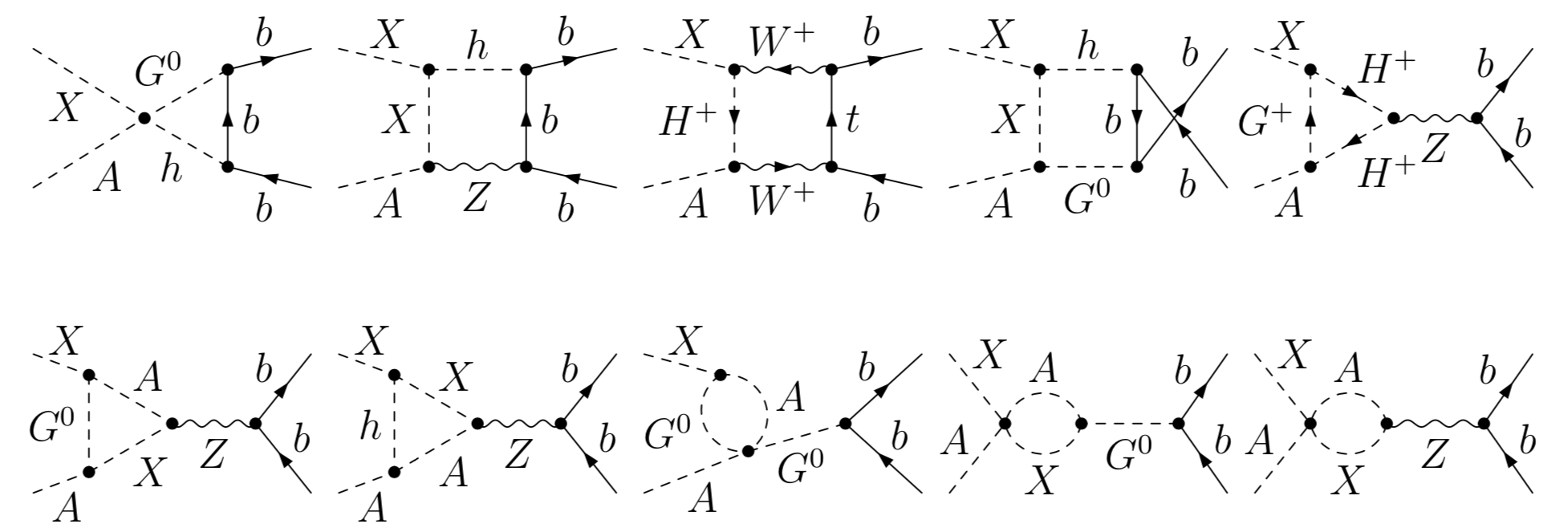}
\caption{\label{fig:coannff-treeaxbb}\it Some one-loop diagrams (in the non-linear Feynman gauge) for $AX \to b \bar b$. We only show a subset of the box corrections and the triangles. Note the last two diagrams involve the quartic coupling within the dark sector, $\l_2$. Unfortunately, the one mediated by the neutral Goldstone is proportional to the fermion mass. The one mediated by the $Z$ will also give a contribution proportional to the fermion mass, since the one-loop integration will give, unlike the tree-level structure involving the difference between the 2 incoming momenta, a new Lorentz structure proportional to the $Z$ 4-momentum which ends up giving a contribution proportional to the final fermion mass. Therefore the $\l_2$ dependence only affects the $S$-wave.}
\end{center}
\end{figure}
Since these co-annihilation processes do not depend on $\l_L$, a fully OS renormalisation exactly along the lines of the renormalisation of the SM can be carried out. The one-loop corrections will have no scale dependence. These processes allow to test whether the use of a running $\alpha$ is a good approximation for the full one-loop corrections. We will show results for all light flavours $f=\nu, l,u,d,b$. For definiteness, we take $\nu=\nu_\tau$, and $l=\mu$. We look at  both $d$ and $b$ since the latter is affected by the heavy top-loop, as is the case for the $Z \to b \bar b$ decay, and also because it may be more sensitive to the $\l_2$ dependence which appears in the $S$-channel. Recall that $\l_2$ measures the self-interaction exclusively within the dark sector, see~\cite{OurPaper1_2020}. We find however that this $\l_2$ dependence is very much suppressed by the mass of the (very) fermions, see Figure~\ref{fig:coannff-treeaxbb}, as we explain below.

\begin{figure}[h]
\begin{center}
\includegraphics[width=0.8\textwidth,height=0.46\textwidth]{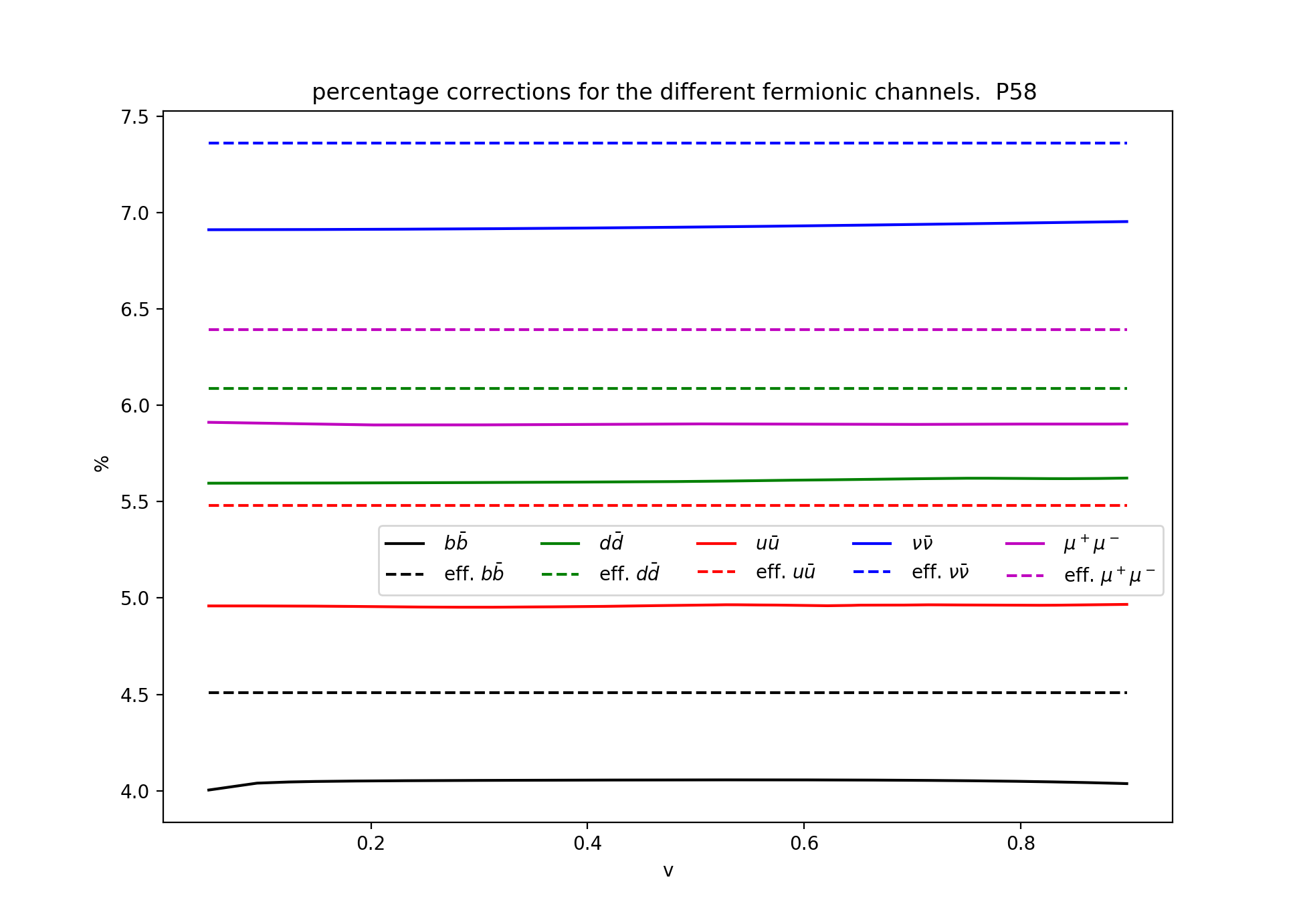}
\caption{\label{fig:loop-coann-allf}\it Full results for all channels. The dashed curves corresponds to the effective cross-sections in Equation~\ref{eq:analytics-coannAXapp}. While a close inspection of the $b \bar b$ channel reveals a very small presence of the $S$-wave contribution at $v \sim 0$, the $\l_2$ dependence is too tiny to be seen on this plot.The $\l_2$ dependence is magnified in Figure~\ref{fig:loopbbarb-l2}.}
\end{center}
\end{figure}

We perform the full one-loop electroweak corrections for both benchmarks P58 and P60. The electroweak corrections, expressed in terms of relative corrections, are practically exactly the same for all channels for both benchmark points.  We therefore only show the results for P58. The computation of the radiative corrections for the charged fermions includes the QED final state radiation that we obtain through the slicing technique, see~\cite{Baro:2007em} for example. For a fermion of charge $Q_f$, these final state QED corrections amount to a relative correction $3 Q_f^2 \alpha/4\pi$, which are very small, $<0.2\%$ (for charged leptons). The QCD final state correction $\sim \alpha_s(M_Z^2)/\pi$ amounts to about $3.6\%$. We have not included these in the figures we show below, Figure~\ref{fig:loop-coann-allf}. 
Apart from the $b$ quarks, the relative corrections for all fermion final states are practically a modest rescaling of the tree-level cross-section. Indeed the relative corrections are velocity independent. For the $b$ quark, this is also the case for all $v$ but $v\sim 0$ where the $S$-wave contributes negligibly. We will come back to this feature. The flavour dependence of the correction is also very small, all corrections being within $2\%$ of each other. Apart from the behaviour at $v\sim 0$, the corrections at all $v$ for the bottom-quark final states are slightly different due to the contribution of the heavy top-quark, a feature known from $Z$-physics. On average, the corrections are about $6\%$ and are therefore a factor 2 lower than if we had used $\alpha(M_Z^2)$. Genuine box and triangle contributions are therefore not negligible. 

Inspired by equation~\ref{eq:analytics-coannAX}, and in view that the $S$-wave contribution is tiny even in the case of $b \bar b$ we suggest to improve the tree-level cross-sections by using  effective cross-sections defined solely in terms of physical observables, 
\beqn
\label{eq:analytics-coannAXapp}
\sigma_{v^2}^{f\bar f,\text{eff.}}&\simeq& \frac{12 \pi}{ s}  \frac{1-m_-^2 m_+^2/s^2}{(1-M_Z^2/s)^2 
}    \frac{\Gamma_{Z \to f \bar f}\; \Gamma_{Z \to \nu \bar \nu}}{M_Z^2}\nonumber \\
\sigma_0^{f \bar f,\text{eff.}}&\simeq&\frac{144 \pi}{ s}  \frac{m_-^2 m_+^2}{s^2-m_-^2 m_+^2} \frac{1}{(1-M_Z^2/s)^2}   \frac{ \Gamma_{Z \to \nu \bar \nu}^2}{M_Z^2}. 
\eeqn
First $g^2/c_W^2$ in~\ref{eq:analytics-coannAX} is traded for $\Gamma(Z\to \nu \bar \nu)$. Then all partial widths in equation~\ref{eq:analytics-coannAXapp} are one-loop corrected partial widths. The results of the almost velocity independent corrections are shown in figure~\ref{fig:loop-coann-allf}. The effective corrective factor misses  only about $-0.5\%$ for the benchmark point P58 (almost independent of the nature of the fermion) of the full one-loop corrections and about $-1\%$ for P60. These small corrections are the effect of the box contributions. These effective approximations are therefore excellent. 

\begin{figure}[hbtp]
\begin{center}
\includegraphics[width=0.48\textwidth,height=0.36\textwidth]{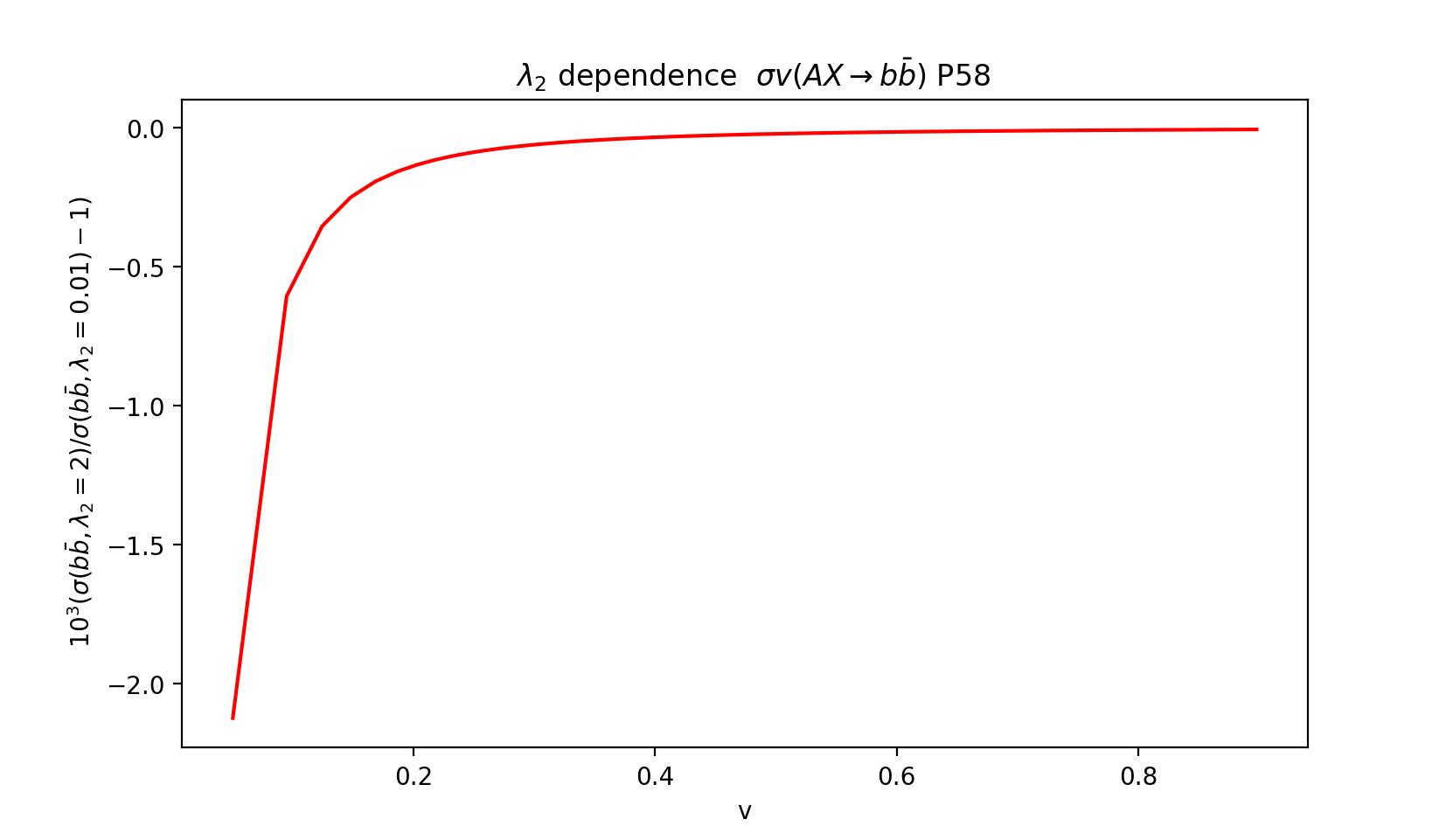}
\caption{\label{fig:loopbbarb-l2}\it The tiny $\l_2$ dependence in the $b\bar b$ co-annihilation channel. The $\l_2$ dependence is measured relative to the $d\bar d$ cross-section where the $\l_2$ dependence is vanishing. Observe that the units in the graph indicate a variation which, at its highest, does not cross a couple of per-mille.}
\end{center}
\end{figure} 

The very tiny contribution of the $S$-wave at very small velocity (where the $P$-wave vanishes) is theoretically interesting even if it is without any phenomenological impact. It could have been larger if the fermion masses were larger. Theoretically, the $S$-wave is sensitive to $\l_2$ as  explained in the caption of Figure~\ref{fig:coannff-treeaxbb}. We verified this property directly in the co-annihilation  into neutrinos where no $\l_2$ dependence was detected numerically and we confirmed that the largest $\l_2$ dependence occurs with the $b$ final state. Unfortunately, the maximum correction from the dark sector is at the level of $2$ per-mille for $v=0$ where the cross-section is smallest. As soon as the $P$-wave contribution kicks in, the tiny $\l_2$ dependence in the $S$-wave contribution is totally swamped such that the one-loop corrected cross-section is, for all purposes,  insensitive to $\l_2$. This very tiny $\l_2$ dependence is shown in Figure~\ref{fig:loopbbarb-l2} where we track the $\l_2$ dependence by considering the ratio in the $b \bar b$ cross-section between $\l_2=0.01$ and $\l_2=2$.


\section{The $XX \to W W^\star$ in the co-annihilation region}
\subsection{Tree-level considerations}
\begin{figure}[hbtp]
\begin{center}
\includegraphics[width=0.8\textwidth,height=0.2\textwidth]{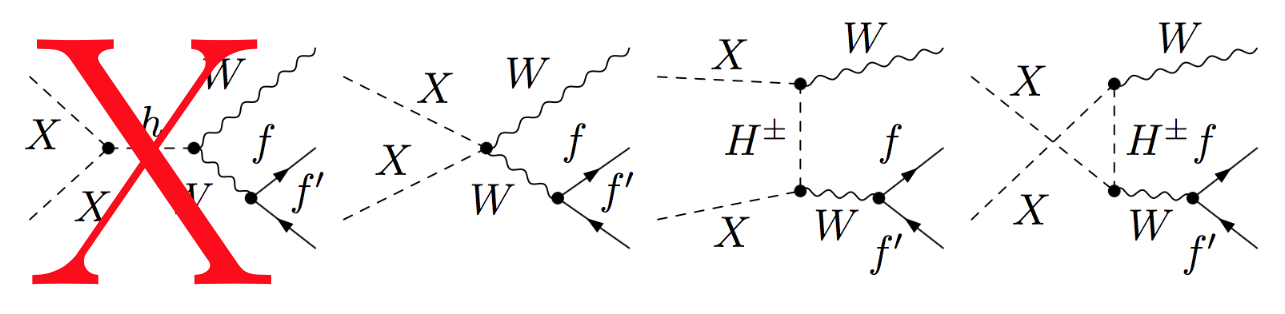}
\caption{\label{fig:FD_loop_wtnt_nohiggs}\it A selection of diagrams for the  tree-level cross-section $XX \to W f \bar f^\prime$. Since $\l_L$=0  for the benchmark point P60, Higgs exchange does not take place for P60.}
\end{center}
\end{figure}

\begin{figure}[htbp]
\begin{center}
\includegraphics[width=0.48\textwidth,height=0.36\textwidth]{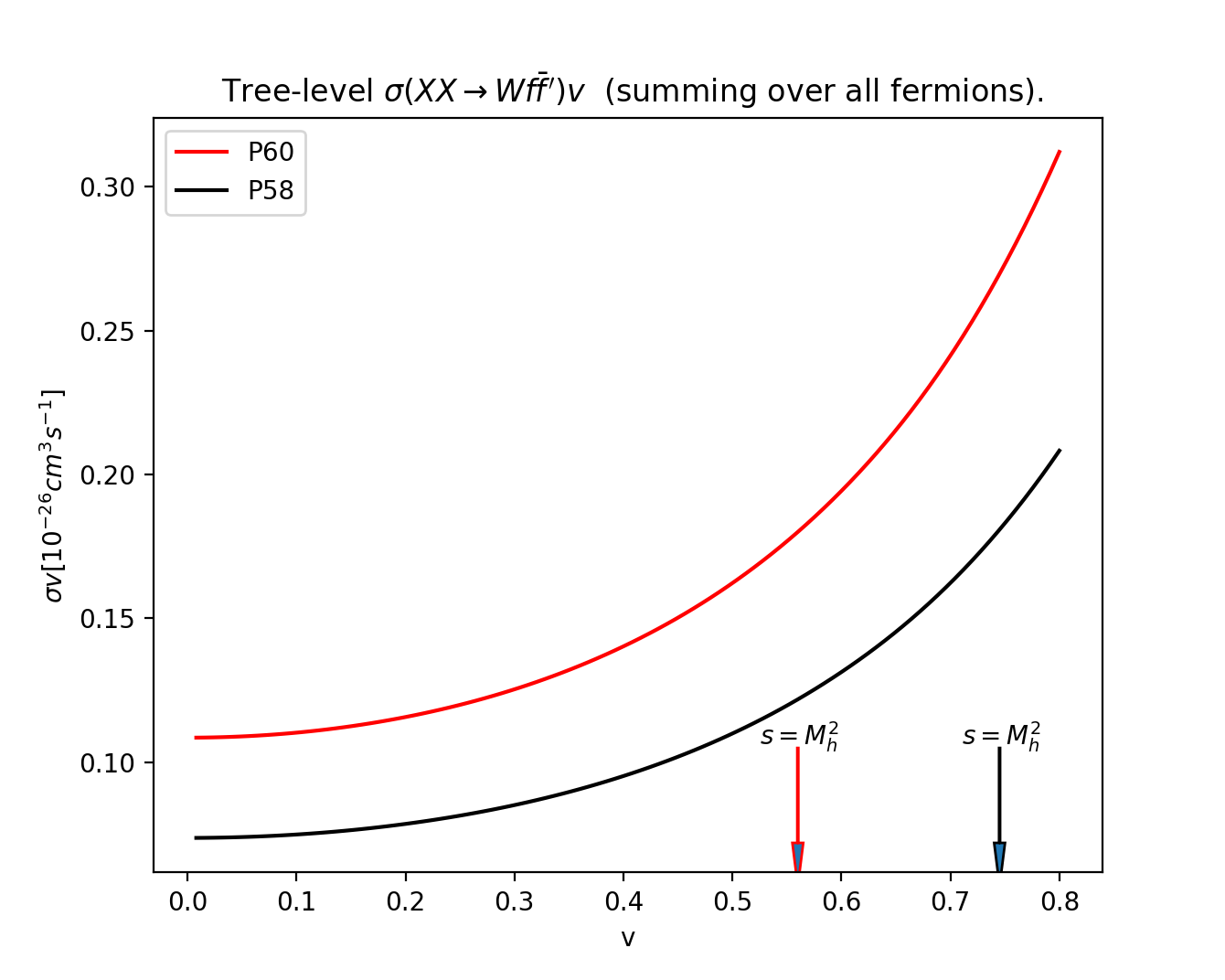}
\caption{\label{fig:tree_wtnt_vP60}\it The tree-level cross-section $XX \to W f \bar f^\prime$ for the benchmark points P58 and P60. The arrow at $v \sim 0.56 (0.745)$ indicates the position corresponding to the Higgs resonance (if crossed) for $M_X=60 \; (58)$ GeV.}
\end{center}
\end{figure}

For P60, $XX$ annihilation to $WW$ below threshold with one $W$ decaying into a fermion pair, accounts at tree-level, for $10\%$ of the relic density. This cross-section does not suffer from the Boltzmann suppression. A non-zero value of $\l_L$ would have added a Higgs exchange contribution as we show in Figure~\ref{fig:FD_loop_wtnt_nohiggs} in the $WW$ channel, and in the $b\bar b$ channel ($XX \xrightarrow{h} b \bar b $). \\
As $v$ increases, the phase space for $WW^\star$ increases and therefore the cross-section increases as Figure~\ref{fig:tree_wtnt_vP60} shows. Note that this cross-section when compared to $AX \to f\bar f$ is far smaller (the latter will be reduced considerably by the Boltzmann factor). This increase with velocity is smooth across the Higgs resonance, $v=0.56$, a threshold that, as expected, is not felt since $\l_L=0$.

\subsection{One-loop results}
\begin{figure}[htbp]
\begin{center}
\includegraphics[width=0.95\textwidth,height=0.4\textwidth]{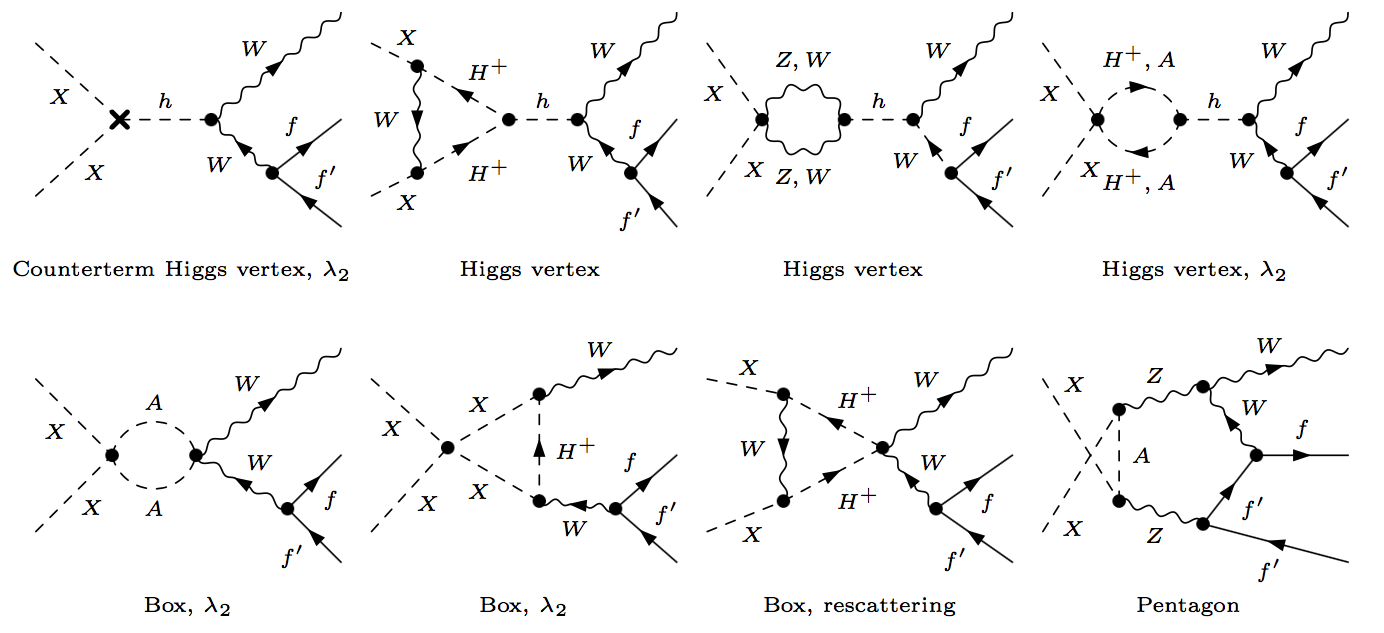}
\caption{\label{fig:FD_loop_wtnt}\it A very small selection of the diagrams that enter the one-loop calculations of $XX \to W W^\star$. We highlight in particular Higgs exchange through the induced $hXX$ vertex at one-loop. We have also picked configurations where a $\l_2$ contribution shows up. These are part of what constitutes rescattering in the dark sector before annihilation into SM particles. An example of a pentagon diagram which can not be considered as a factorised $XX \to WW^\star \to W f f^\prime$ is also singled out (5-point function).}
\end{center}
\end{figure}
\begin{figure}[hbtp]
\begin{center}
\includegraphics[width=0.7\textwidth,height=0.5\textwidth]{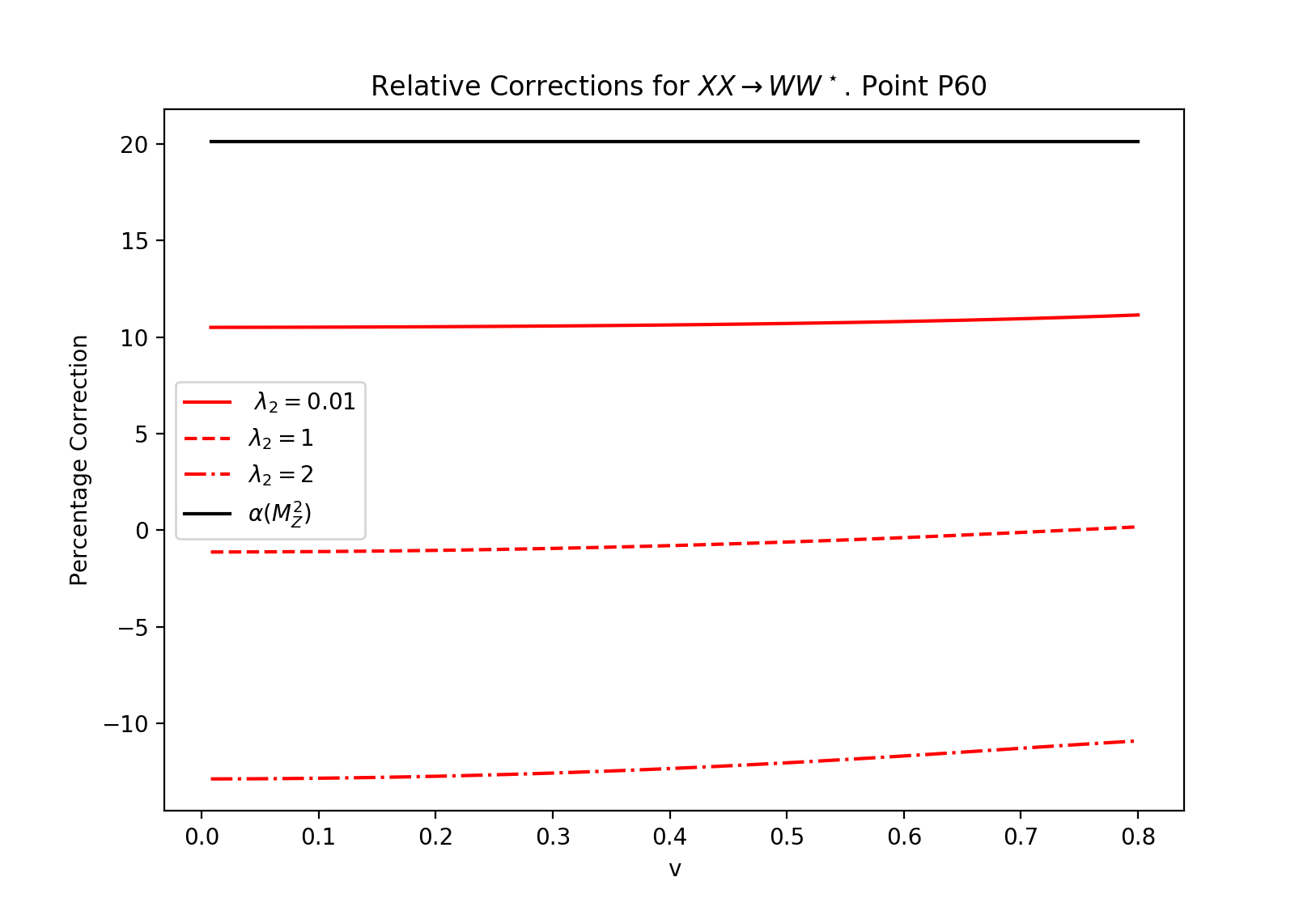}
\caption{\label{fig:coann_loop_l2_wtnt}\it The relative one-loop contribution to $XX \to W f f^\prime$ as a function of the relative velocity $v$ for $\l_2=0.01,1,2$. These corrections are compared to the effect of using an effective tree-level correction with $\alpha(M_Z^2)$.}
\end{center}
\end{figure}

While $\l_L$=0 at tree-level for $XX \to W f f^\prime$ with no SM Higgs exchange contributing, a renormalisation of $\l_L$ is still called for in this process. This is because in {\em general} the $XX \to WW$ amplitude does depend {\em parametrically} on $\l_L$. As Figure~\ref{fig:FD_loop_wtnt} clearly shows, an induced one-loop $hXX$ is generated which entails a one-loop Higgs exchange contribution. The critical question is whether this will lead to an instability when the pole at $s=M_h^2$ is reached? Do we then have to, and how to,  include a width? casting doubt on the organisation of the perturbation series? This same crucial point is dealt with in detail when we study cases where the resonance is present already at ``tree-level". In the present case, $\l_L=0$, there is no need to include a width. Since $h \to XX$ is open, the OS scheme is most appropriate for a definition and renormalisation of $\l_L$. The OS scheme means that we trade $\l_L$ with $\Gamma_{h\to XX}$. In this particular case, the input translates into $\Gamma_{h \to XX}=0$. But $\Gamma_{h \to XX}=0$ is maintained at \underline{all} orders entails, in fact, that the renormalised amplitude for ${{\cal M}}_{h\to XX}(s)=0$ at the renormalisation point $s=M_h^2$, $s$ is the invariant mass of the $XX$ system. For a selection of one-loop contributions that make the $h \to XX$ vertex amplitude see Figure~\ref{fig:FD_loop_wtnt}. 
Consequently, ${{\cal M}}_{h \to XX}(s) \frac{1}{s-M_h^2}|_{s=M_h^2}$ is continuous and the Laurent's series is well defined without a regulator, the width. There is then no pole structure as such and away from $s=M_h^2$ other (non-resonant) structures in the amplitude contribute as importantly. What is very interesting is that the one-loop amplitude is now $\l_2$ dependent, see Figure~\ref{fig:FD_loop_wtnt}. Note that $XX \to W W^\star$ is an abuse of language since such an off-shell amplitude is not an element of the S-matrix. At one-loop, this statement is supported by the appearance of essential pentagon contributions where a split into a $WW^\star$ part is not possible. Nonetheless, the fact that the final state fermions are massless, we find no flavour dependence in the normalised (with respect to the tree-level) loop corrections when the full one-loop contribution is taken into account. {\it In fine}, this is reminiscent and strongly related to the fact that (normalised) electroweak corrections to $W$ decay into fermion pairs are flavour independent~\cite{Bardin:1986fi}. Our results for the electroweak corrections in this channel are displayed in Figure~\ref{fig:coann_loop_l2_wtnt}. The behaviour with respect to velocity is very smooth, for all values of $\l_2$, confirming that the location of the Higgs is crossed continuously. 
The $\l_2$ dependence at one-loop is important. The full one-loop correction is about $10\%$ for the smallest of $\l_2$, $\l_2=0.01$, it gives an almost vanishing total correction for $\l_2=1$ and then decreases to about $-10\%$ for $\l_2=2$. For this value, $\l_2=2$, there is a $30\%$ difference with the use of an effective $\alpha(M_Z^2)$. Therefore, $\alpha(M_Z^2)$ does not perform well as an approximation. The velocity dependence of the $\l_2$ correction is very weak. To summarise, at the level of the cross-sections the $\l_2$ contribution is important and discriminating. Will this conclusion still hold when we call these cross-sections to compute the relic density considering that the weight of this channel is only $10\%$?


\section{Effect on the relic density}
\begin{table}[hbt]\begin{center}
\begin{tabular}{cl|l|c|c||}
\cline{3-5}
 &  &$\Omega h^2$  & $AX \to f \bar f$ & $WW^*$  \\
\cline{3-5} \hline 
\multicolumn{1}{ ||c  }{\multirow{4}{*}{{\bf P58}}}& \multicolumn{1}{ |l| }{tree}& 0.113 & 95\% & 5\% \\
\multicolumn{1}{ ||c  }{} & \multicolumn{1}{ |l| }{ One-loop $AX \to f\bar f$ and  tree $XX \to WW^\star$} & 0.108 (-4.05\%) & 95\% & 5\% \\
\multicolumn{1}{ ||c  }{} &\multicolumn{1}{ |l| }{Effective $AX \to f\bar f$ and  tree $XX \to WW^\star$} & 0.108 (-4.45\%) & 95\% &5\%\\
 \multicolumn{1}{ ||c  }{} & \multicolumn{1}{ |l| }{$\alpha(M_Z^2)$}& 0.101 (-10.62\%) & 95\% & 5\% \\
 \hline \hline 
\multicolumn{1}{ ||c  }{\multirow{9}{*}{{\bf P60}}}& \multicolumn{1}{ |l| }{tree}& 0.116 & 91\% & 9\% \\
\multicolumn{1}{ ||c  }{} & \multicolumn{1}{ |l| }{ Full one-loop, $\l_2=0.01$}& 0.111 (-4.04\%) & 91\% & 9\% \\
\multicolumn{1}{ ||c  }{} & \multicolumn{1}{ |l| }{ Full one-loop, $\l_2=1$}& 0.112 (-3.43\%) & 91\% & 9\% \\
\multicolumn{1}{ ||c  }{} & \multicolumn{1}{ |l| }{ Full one-loop, $\l_2=2$}& 0.113 (-2.81\%) & 91\% & 9\% \\
\multicolumn{1}{ ||c  }{} &\multicolumn{1}{ |l| }{ Effective $AX \to f\bar f$ and  tree $XX \to WW^\star$} & 0.111 (-4.36\%) & 91\% &8\% \\
\multicolumn{1}{ ||c  }{} &\multicolumn{1}{ |l| }{ Effective $AX \to f\bar f$ and loop $XX \to WW^\star, \l_2=0.01$} & 0.110 (-4.92\%) & 91\% &9\% \\
\multicolumn{1}{ ||c  }{} &\multicolumn{1}{ |l| }{ Effective $AX \to f\bar f$ and  loop $XX \to WW^\star, \l_2=1$} & 0.111 (-4.32\%) & 91\% &9\% \\
\multicolumn{1}{ ||c  }{} &\multicolumn{1}{ |l| }{ Effective $AX \to f\bar f$ and  loop $XX \to WW^\star, \l_2=2$} & 0.112 (-3.71\%) & 91\% &9\% \\
 \multicolumn{1}{ ||c  }{}& \multicolumn{1}{ |l| }{ $\alpha(M_Z^2)$ }& 0.103 (-11.21\%) & 91\% & 9\% \\
\hline
\end{tabular}
\end{center}
\caption{\it The relic density for P58 and P60 with cross-sections computed at tree-level and after including the full electroweak corrections, as well as implementing the effective cross-sections for $AX \to f \bar f$ and the cross-sections calculated with $\alpha(M_Z^2)$. The percentage correction to the relic density is given in parenthesis. The percentage weight of each channel is computed for each implementation of the corrected cross-section.}
\label{tab:rel-cont-58-60}
\end{table}
In order to translate the cross-sections into a prediction on the relic density, we interface our calculations of the one-loop corrected cross-section with {\tt microMEGAs}~\cite{Belanger:2001fz, Belanger:2004yn, Belanger:2006is, Belanger:2013oya, Belanger:2018mqt} to perform the convolution of the velocity and temperature and solving the Boltzmann equation. Table~\ref{tab:rel-cont-58-60} shows that the correction to the relic density for these low mass co-annihilation configurations are small with a correction in line with what we found for the dominant cross-sections $AX \to f \bar f$ (remember that $\Omega \propto 1/\sigma$) which are practically $v$-independent. As expected, the corrections are much smaller than those found with the approximation of using a running $\alpha$. Since in P58 the contribution of the $WW^\star$ channel is less than about $5\%$, we do not include the effect of the electroweak corrections in this channel. However, we include these corrections for P60 since this channel accounts for almost $10\%$ of the total relic density (at tree-level). While any small $\l_2$ dependence (which occurs at the very small $v$ in $AX\to f \bar f$) is washed out when the corrections are translated into the relic density calculation, the $\l_2$ dependence observed in the cross-sections $XX \to W f \bar f^\prime$, make their way into the relic density even if the $WW^\star$ channel account for only $\sim 10\%$. For $\l_2=2$, the correction in the $WW^\star$ channel counterbalances part of the correction from the contributions $AX \to f \bar f$ leading to an overall correction on the relic density less than $3\%$. As a summary, in the co-annihilation region, if one is satisfied with a calculation of the relic density with an accuracy no better than $5\%$, one should content oneself with a tree-level calculation. An ``improved" ($\alpha (M_Z^2)$) gives a correction of about $10\%$. The effective cross-sections that are based on re-expressing the $AX \to f \bar f^\prime$ in terms of the one-loop partial decay widths of the $Z\to f\bar f$ reproduce the full one-loop results within $1\%$.


\section{Conclusions}
We have studied two IDM low mass DM benchmark points where co-annihilation is the dominant contribution to the relic density. The co-annihilation into fermions is driven essentially by the SM gauge coupling. The relative corrections are small. We show that they can be parameterised by simple effective cross-sections involving $Z$-observables. The fermion flavour dependence is small. The small contribution from $W W^\star$ to the relic density is interesting since at one-loop it reveals an important $\l_2$ dependence that accesses the purely dark sector. However, this $\l_2$ dependence gets diluted because of the small weight of $W W^\star$ to the relic density. Nonetheless, our calculations show that performing a full one-loop calculation gives smaller corrections than using an effective $\alpha (M_Z^2)$ into tree-level cross-sections. 

\acknowledgments
We thank Alexander Pukhov for  helpful discussions. HS is supported by the National Natural Science Foundation of China (Grant No.12075043, No.11675033). He warmly thanks the CPTGA and LAPTh for support during his visit to France in 2019 when this work was initiated. SB is grateful for the support received from IPPP, Durham, UK, where most of this work was performed and is also grateful to the support received from LAPTh, in 2019, where this work was initiated. NC is financially supported by IISc (Indian Institute of Science, Bangalore, India) through the C.V.~Raman postdoctoral fellowship. He also acknowledges the support received from DST, India, under grant number IFA19-PH237 (INSPIRE Faculty Award).

\bibliography{../../NLO_IDM570}

\providecommand{\href}[2]{#2}\begingroup\raggedright\begin{thebibliography}{10}

\bibitem{OurPaper1_2020}
S.~Banerjee, F.~Boudjema, N.~Chakrabarty and H.~Sun, \emph{{Relic density of
  dark matter in the inert doublet model beyond leading order: Renormalisation
  and Constraints for the low mass region}}, .

\bibitem{Deshpande:1977rw}
N.~G. Deshpande and E.~Ma, \emph{{Pattern of Symmetry Breaking with Two Higgs
  Doublets}}, \href{http://dx.doi.org/10.1103/PhysRevD.18.2574}{\emph{Phys.
  Rev.} {\bf D18} (1978) 2574}.

\bibitem{Barbieri:2006dq}
R.~Barbieri, L.~J. Hall and V.~S. Rychkov, \emph{{Improved naturalness with a
  heavy Higgs: An Alternative road to LHC physics}},
  \href{http://dx.doi.org/10.1103/PhysRevD.74.015007}{\emph{Phys. Rev.} {\bf
  D74} (2006) 015007}, [\href{http://arxiv.org/abs/hep-ph/0603188}{{\tt
  hep-ph/0603188}}].

\bibitem{Hambye:2007vf}
T.~Hambye and M.~H.~G. Tytgat, \emph{{Electroweak symmetry breaking induced by
  dark matter}},
  \href{http://dx.doi.org/10.1016/j.physletb.2007.11.069}{\emph{Phys. Lett.}
  {\bf B659} (2008) 651--655}, [\href{http://arxiv.org/abs/0707.0633}{{\tt
  0707.0633}}].

\bibitem{LopezHonorez:2006gr}
L.~Lopez~Honorez, E.~Nezri, J.~F. Oliver and M.~H.~G. Tytgat, \emph{{The Inert
  Doublet Model: An Archetype for Dark Matter}},
  \href{http://dx.doi.org/10.1088/1475-7516/2007/02/028}{\emph{JCAP} {\bf 0702}
  (2007) 028}, [\href{http://arxiv.org/abs/hep-ph/0612275}{{\tt
  hep-ph/0612275}}].

\bibitem{Cao:2007rm}
Q.-H. Cao, E.~Ma and G.~Rajasekaran, \emph{{Observing the Dark Scalar Doublet
  and its Impact on the Standard-Model Higgs Boson at Colliders}},
  \href{http://dx.doi.org/10.1103/PhysRevD.76.095011}{\emph{Phys. Rev.} {\bf
  D76} (2007) 095011}, [\href{http://arxiv.org/abs/0708.2939}{{\tt
  0708.2939}}].

\bibitem{Gustafsson:2007pc}
M.~Gustafsson, E.~Lundstrom, L.~Bergstrom and J.~Edsjo, \emph{{Significant
  Gamma Lines from Inert Higgs Dark Matter}},
  \href{http://dx.doi.org/10.1103/PhysRevLett.99.041301}{\emph{Phys. Rev.
  Lett.} {\bf 99} (2007) 041301},
  [\href{http://arxiv.org/abs/astro-ph/0703512}{{\tt astro-ph/0703512}}].

\bibitem{Agrawal:2008xz}
P.~Agrawal, E.~M. Dolle and C.~A. Krenke, \emph{{Signals of Inert Doublet Dark
  Matter in Neutrino Telescopes}},
  \href{http://dx.doi.org/10.1103/PhysRevD.79.015015}{\emph{Phys. Rev.} {\bf
  D79} (2009) 015015}, [\href{http://arxiv.org/abs/0811.1798}{{\tt
  0811.1798}}].

\bibitem{Hambye:2009pw}
T.~Hambye, F.~S. Ling, L.~Lopez~Honorez and J.~Rocher, \emph{{Scalar Multiplet
  Dark Matter}}, \href{http://dx.doi.org/10.1007/JHEP05(2010)066,
  10.1088/1126-6708/2009/07/090}{\emph{JHEP} {\bf 07} (2009) 090},
  [\href{http://arxiv.org/abs/0903.4010}{{\tt 0903.4010}}].

\bibitem{Lundstrom:2008ai}
E.~Lundstrom, M.~Gustafsson and J.~Edsjo, \emph{{The Inert Doublet Model and
  LEP II Limits}},
  \href{http://dx.doi.org/10.1103/PhysRevD.79.035013}{\emph{Phys. Rev.} {\bf
  D79} (2009) 035013}, [\href{http://arxiv.org/abs/0810.3924}{{\tt
  0810.3924}}].

\bibitem{Andreas:2009hj}
S.~Andreas, M.~H.~G. Tytgat and Q.~Swillens, \emph{{Neutrinos from Inert
  Doublet Dark Matter}},
  \href{http://dx.doi.org/10.1088/1475-7516/2009/04/004}{\emph{JCAP} {\bf 0904}
  (2009) 004}, [\href{http://arxiv.org/abs/0901.1750}{{\tt 0901.1750}}].

\bibitem{Arina:2009um}
C.~Arina, F.-S. Ling and M.~H.~G. Tytgat, \emph{{IDM and iDM or The Inert
  Doublet Model and Inelastic Dark Matter}},
  \href{http://dx.doi.org/10.1088/1475-7516/2009/10/018}{\emph{JCAP} {\bf 0910}
  (2009) 018}, [\href{http://arxiv.org/abs/0907.0430}{{\tt 0907.0430}}].

\bibitem{Dolle:2009ft}
E.~Dolle, X.~Miao, S.~Su and B.~Thomas, \emph{{Dilepton Signals in the Inert
  Doublet Model}},
  \href{http://dx.doi.org/10.1103/PhysRevD.81.035003}{\emph{Phys. Rev.} {\bf
  D81} (2010) 035003}, [\href{http://arxiv.org/abs/0909.3094}{{\tt
  0909.3094}}].

\bibitem{Nezri:2009jd}
E.~Nezri, M.~H.~G. Tytgat and G.~Vertongen, \emph{{e+ and anti-p from inert
  doublet model dark matter}},
  \href{http://dx.doi.org/10.1088/1475-7516/2009/04/014}{\emph{JCAP} {\bf 0904}
  (2009) 014}, [\href{http://arxiv.org/abs/0901.2556}{{\tt 0901.2556}}].

\bibitem{Miao:2010rg}
X.~Miao, S.~Su and B.~Thomas, \emph{{Trilepton Signals in the Inert Doublet
  Model}}, \href{http://dx.doi.org/10.1103/PhysRevD.82.035009}{\emph{Phys.
  Rev.} {\bf D82} (2010) 035009}, [\href{http://arxiv.org/abs/1005.0090}{{\tt
  1005.0090}}].

\bibitem{Gong:2012ri}
J.-O. Gong, H.~M. Lee and S.~K. Kang, \emph{{Inflation and dark matter in two
  Higgs doublet models}},
  \href{http://dx.doi.org/10.1007/JHEP04(2012)128}{\emph{JHEP} {\bf 04} (2012)
  128}, [\href{http://arxiv.org/abs/1202.0288}{{\tt 1202.0288}}].

\bibitem{Gustafsson:2012aj}
M.~Gustafsson, S.~Rydbeck, L.~Lopez-Honorez and E.~Lundstrom, \emph{{Status of
  the Inert Doublet Model and the Role of multileptons at the LHC}},
  \href{http://dx.doi.org/10.1103/PhysRevD.86.075019}{\emph{Phys. Rev.} {\bf
  D86} (2012) 075019}, [\href{http://arxiv.org/abs/1206.6316}{{\tt
  1206.6316}}].

\bibitem{Swiezewska:2012eh}
B.~Swiezewska and M.~Krawczyk, \emph{{Diphoton rate in the inert doublet model
  with a 125 GeV Higgs boson}},
  \href{http://dx.doi.org/10.1103/PhysRevD.88.035019}{\emph{Phys. Rev.} {\bf
  D88} (2013) 035019}, [\href{http://arxiv.org/abs/1212.4100}{{\tt
  1212.4100}}].

\bibitem{Arhrib:2012ia}
A.~Arhrib, R.~Benbrik and N.~Gaur, \emph{{$H\to \gamma \gamma$ in Inert Higgs
  Doublet Model}},
  \href{http://dx.doi.org/10.1103/PhysRevD.85.095021}{\emph{Phys. Rev.} {\bf
  D85} (2012) 095021}, [\href{http://arxiv.org/abs/1201.2644}{{\tt
  1201.2644}}].

\bibitem{Wang:2012zv}
L.~Wang and X.-F. Han, \emph{{LHC diphoton Higgs signal and top quark
  forward-backward asymmetry in quasi-inert Higgs doublet model}},
  \href{http://dx.doi.org/10.1007/JHEP05(2012)088}{\emph{JHEP} {\bf 05} (2012)
  088}, [\href{http://arxiv.org/abs/1203.4477}{{\tt 1203.4477}}].

\bibitem{Goudelis:2013uca}
A.~Goudelis, B.~Herrmann and O.~Stal, \emph{{Dark matter in the Inert Doublet
  Model after the discovery of a Higgs-like boson at the LHC}},
  \href{http://dx.doi.org/10.1007/JHEP09(2013)106}{\emph{JHEP} {\bf 09} (2013)
  106}, [\href{http://arxiv.org/abs/1303.3010}{{\tt 1303.3010}}].

\bibitem{Arhrib:2013ela}
A.~Arhrib, Y.-L.~S. Tsai, Q.~Yuan and T.-C. Yuan, \emph{{An Updated Analysis of
  Inert Higgs Doublet Model in light of the Recent Results from LUX, PLANCK,
  AMS-02 and LHC}},
  \href{http://dx.doi.org/10.1088/1475-7516/2014/06/030}{\emph{JCAP} {\bf 1406}
  (2014) 030}, [\href{http://arxiv.org/abs/1310.0358}{{\tt 1310.0358}}].

\bibitem{Krawczyk:2013jta}
M.~Krawczyk, D.~Sokolowska, P.~Swaczyna and B.~Swiezewska, \emph{{Constraining
  Inert Dark Matter by $R_{\gamma\gamma}$ and WMAP data}},
  \href{http://dx.doi.org/10.1007/JHEP09(2013)055}{\emph{JHEP} {\bf 09} (2013)
  055}, [\href{http://arxiv.org/abs/1305.6266}{{\tt 1305.6266}}].

\bibitem{Osland:2013sla}
P.~Osland, A.~Pukhov, G.~M. Pruna and M.~Purmohammadi, \emph{{Phenomenology of
  charged scalars in the CP-Violating Inert-Doublet Model}},
  \href{http://dx.doi.org/10.1007/JHEP04(2013)040}{\emph{JHEP} {\bf 04} (2013)
  040}, [\href{http://arxiv.org/abs/1302.3713}{{\tt 1302.3713}}].

\bibitem{Abe:2015rja}
T.~Abe and R.~Sato, \emph{{Quantum corrections to the spin-independent cross
  section of the inert doublet dark matter}},
  \href{http://dx.doi.org/10.1007/JHEP03(2015)109}{\emph{JHEP} {\bf 03} (2015)
  109}, [\href{http://arxiv.org/abs/1501.04161}{{\tt 1501.04161}}].

\bibitem{Blinov:2015qva}
N.~Blinov, J.~Kozaczuk, D.~E. Morrissey and A.~de~la Puente, \emph{{Compressing
  the Inert Doublet Model}},
  \href{http://dx.doi.org/10.1103/PhysRevD.93.035020}{\emph{Phys. Rev.} {\bf
  D93} (2016) 035020}, [\href{http://arxiv.org/abs/1510.08069}{{\tt
  1510.08069}}].

\bibitem{Diaz:2015pyv}
M.~A. Díaz, B.~Koch and S.~Urrutia-Quiroga, \emph{{Constraints to Dark Matter
  from Inert Higgs Doublet Model}},
  \href{http://dx.doi.org/10.1155/2016/8278375}{\emph{Adv. High Energy Phys.}
  {\bf 2016} (2016) 8278375}, [\href{http://arxiv.org/abs/1511.04429}{{\tt
  1511.04429}}].

\bibitem{Ilnicka:2015jba}
A.~Ilnicka, M.~Krawczyk and T.~Robens, \emph{{Inert Doublet Model in light of
  LHC Run I and astrophysical data}},
  \href{http://dx.doi.org/10.1103/PhysRevD.93.055026}{\emph{Phys. Rev.} {\bf
  D93} (2016) 055026}, [\href{http://arxiv.org/abs/1508.01671}{{\tt
  1508.01671}}].

\bibitem{Belanger:2015kga}
G.~Belanger, B.~Dumont, A.~Goudelis, B.~Herrmann, S.~Kraml and D.~Sengupta,
  \emph{{Dilepton constraints in the Inert Doublet Model from Run1 of the
  LHC}}, \href{http://dx.doi.org/10.1103/PhysRevD.91.115011}{\emph{Phys. Rev.}
  {\bf D91} (2015) 115011}, [\href{http://arxiv.org/abs/1503.07367}{{\tt
  1503.07367}}].

\bibitem{Carmona:2015haa}
A.~Carmona and M.~Chala, \emph{{Composite Dark Sectors}},
  \href{http://dx.doi.org/10.1007/JHEP06(2015)105}{\emph{JHEP} {\bf 06} (2015)
  105}, [\href{http://arxiv.org/abs/1504.00332}{{\tt 1504.00332}}].

\bibitem{Kanemura:2016sos}
S.~Kanemura, M.~Kikuchi and K.~Sakurai, \emph{{Testing the dark matter scenario
  in the inert doublet model by future precision measurements of the Higgs
  boson couplings}},
  \href{http://dx.doi.org/10.1103/PhysRevD.94.115011}{\emph{Phys. Rev.} {\bf
  D94} (2016) 115011}, [\href{http://arxiv.org/abs/1605.08520}{{\tt
  1605.08520}}].

\bibitem{Queiroz:2015utg}
F.~S. Queiroz and C.~E. Yaguna, \emph{{The CTA aims at the Inert Doublet
  Model}}, \href{http://dx.doi.org/10.1088/1475-7516/2016/02/038}{\emph{JCAP}
  {\bf 1602} (2016) 038}, [\href{http://arxiv.org/abs/1511.05967}{{\tt
  1511.05967}}].

\bibitem{Belyaev:2016lok}
A.~Belyaev, G.~Cacciapaglia, I.~P. Ivanov, F.~Rojas-Abatte and M.~Thomas,
  \emph{{Anatomy of the Inert Two Higgs Doublet Model in the light of the LHC
  and non-LHC Dark Matter Searches}},
  \href{http://dx.doi.org/10.1103/PhysRevD.97.035011}{\emph{Phys. Rev.} {\bf
  D97} (2018) 035011}, [\href{http://arxiv.org/abs/1612.00511}{{\tt
  1612.00511}}].

\bibitem{Arcadi:2019lka}
G.~Arcadi, A.~Djouadi and M.~Raidal, \emph{{Dark Matter through the Higgs
  portal}}, \href{http://dx.doi.org/10.1016/j.physrep.2019.11.003}{\emph{Phys.
  Rept.} {\bf 842} (2020) 1--180}, [\href{http://arxiv.org/abs/1903.03616}{{\tt
  1903.03616}}].

\bibitem{Eiteneuer:2017hoh}
B.~Eiteneuer, A.~Goudelis and J.~Heisig, \emph{{The inert doublet model in the
  light of Fermi-LAT gamma-ray data: a global fit analysis}},
  \href{http://dx.doi.org/10.1140/epjc/s10052-017-5166-1}{\emph{Eur. Phys. J.}
  {\bf C77} (2017) 624}, [\href{http://arxiv.org/abs/1705.01458}{{\tt
  1705.01458}}].

\bibitem{Ilnicka:2018def}
A.~Ilnicka, T.~Robens and T.~Stefaniak, \emph{{Constraining Extended Scalar
  Sectors at the LHC and beyond}},
  \href{http://dx.doi.org/10.1142/S0217732318300070}{\emph{Mod. Phys. Lett.}
  {\bf A33} (2018) 1830007}, [\href{http://arxiv.org/abs/1803.03594}{{\tt
  1803.03594}}].

\bibitem{Kalinowski:2018ylg}
J.~Kalinowski, W.~Kotlarski, T.~Robens, D.~Sokolowska and A.~F. Zarnecki,
  \emph{{Benchmarking the Inert Doublet Model for $e^+ e^-$ colliders}},
  \href{http://dx.doi.org/10.1007/JHEP12(2018)081}{\emph{JHEP} {\bf 12} (2018)
  081}, [\href{http://arxiv.org/abs/1809.07712}{{\tt 1809.07712}}].

\bibitem{Ferreira:2009jb}
P.~M. Ferreira and D.~R.~T. Jones, \emph{{Bounds on scalar masses in two Higgs
  doublet models}},
  \href{http://dx.doi.org/10.1088/1126-6708/2009/08/069}{\emph{JHEP} {\bf 08}
  (2009) 069}, [\href{http://arxiv.org/abs/0903.2856}{{\tt 0903.2856}}].

\bibitem{Ferreira:2015pfi}
P.~M. Ferreira and B.~Swiezewska, \emph{{One-loop contributions to neutral
  minima in the inert doublet model}},
  \href{http://dx.doi.org/10.1007/JHEP04(2016)099}{\emph{JHEP} {\bf 04} (2016)
  099}, [\href{http://arxiv.org/abs/1511.02879}{{\tt 1511.02879}}].

\bibitem{Kanemura:2002vm}
S.~Kanemura, S.~Kiyoura, Y.~Okada, E.~Senaha and C.~P. Yuan, \emph{{New physics
  effect on the Higgs selfcoupling}},
  \href{http://dx.doi.org/10.1016/S0370-2693(03)00268-5}{\emph{Phys. Lett.}
  {\bf B558} (2003) 157--164}, [\href{http://arxiv.org/abs/hep-ph/0211308}{{\tt
  hep-ph/0211308}}].

\bibitem{Senaha:2018xek}
E.~Senaha, \emph{{Radiative Corrections to Triple Higgs Coupling and
  Electroweak Phase Transition: Beyond One-loop Analysis}},
  \href{http://dx.doi.org/10.1103/PhysRevD.100.055034}{\emph{Phys. Rev. D} {\bf
  100} (2019) 055034}, [\href{http://arxiv.org/abs/1811.00336}{{\tt
  1811.00336}}].

\bibitem{Braathen:2019pxr}
J.~Braathen and S.~Kanemura, \emph{{On two-loop corrections to the Higgs
  trilinear coupling in models with extended scalar sectors}},
  \href{http://dx.doi.org/10.1016/j.physletb.2019.07.021}{\emph{Phys. Lett. B}
  {\bf 796} (2019) 38--46}, [\href{http://arxiv.org/abs/1903.05417}{{\tt
  1903.05417}}].

\bibitem{Arhrib:2015hoa}
A.~Arhrib, R.~Benbrik, J.~El~Falaki and A.~Jueid, \emph{{Radiative corrections
  to the Triple Higgs Coupling in the Inert Higgs Doublet Model}},
  \href{http://dx.doi.org/10.1007/JHEP12(2015)007}{\emph{JHEP} {\bf 12} (2015)
  007}, [\href{http://arxiv.org/abs/1507.03630}{{\tt 1507.03630}}].

\bibitem{Garcia-Cely:2015khw}
C.~Garcia-Cely, M.~Gustafsson and A.~Ibarra, \emph{{Probing the Inert Doublet
  Dark Matter Model with Cherenkov Telescopes}},
  \href{http://dx.doi.org/10.1088/1475-7516/2016/02/043}{\emph{JCAP} {\bf 1602}
  (2016) 043}, [\href{http://arxiv.org/abs/1512.02801}{{\tt 1512.02801}}].

\bibitem{Banerjee:2016vrp}
S.~Banerjee and N.~Chakrabarty, \emph{{A revisit to scalar dark matter with
  radiative corrections}},
  \href{http://dx.doi.org/10.1007/JHEP05(2019)150}{\emph{JHEP} {\bf 05} (2019)
  150}, [\href{http://arxiv.org/abs/1612.01973}{{\tt 1612.01973}}].

\bibitem{Basu:2020qoe}
R.~Basu, S.~Banerjee, M.~Pandey and D.~Majumdar, \emph{{Lower bounds on dark
  matter annihilation cross-sections by studying the fluctuations of 21-cm line
  with dark matter candidate in inert doublet model (IDM) with the combined
  effects of dark matter scattering and annihilation}},
  \href{http://arxiv.org/abs/2010.11007}{{\tt 2010.11007}}.

\bibitem{Abouabid:2020eik}
H.~Abouabid, A.~Arhrib, R.~Benbrik, J.~E. Falaki, B.~Gong, W.~Xie et~al.,
  \emph{{One-loop radiative corrections to $e^+ e^-\to Zh^0/H^0A^0$ in the
  Inert Higgs Doublet Model}},  \href{http://arxiv.org/abs/2009.03250}{{\tt
  2009.03250}}.

\bibitem{Kalinowski:2020rmb}
J.~Kalinowski, T.~Robens, D.~Sokolowska and A.~F. Zarnecki, \emph{{IDM
  benchmarks for the LHC and future colliders}},
  \href{http://arxiv.org/abs/2012.14818}{{\tt 2012.14818}}.

\bibitem{Datta:2016nfz}
A.~Datta, N.~Ganguly, N.~Khan and S.~Rakshit, \emph{{Exploring collider
  signatures of the inert Higgs doublet model}},
  \href{http://dx.doi.org/10.1103/PhysRevD.95.015017}{\emph{Phys. Rev.} {\bf
  D95} (2017) 015017}, [\href{http://arxiv.org/abs/1610.00648}{{\tt
  1610.00648}}].

\bibitem{Ade:2015xua}
{\scshape Planck} collaboration, P.~A.~R. Ade et~al., \emph{{Planck 2015
  results. XIII. Cosmological parameters}},
  \href{http://dx.doi.org/10.1051/0004-6361/201525830}{\emph{Astron.
  Astrophys.} {\bf 594} (2016) A13},
  [\href{http://arxiv.org/abs/1502.01589}{{\tt 1502.01589}}].

\bibitem{OurPaper3_2020}
S.~Banerjee, F.~Boudjema, N.~Chakrabarty and H.~Sun, \emph{{Relic density of
  dark matter in the inert doublet model beyond leading order: Annihilation in
  3-body final state for the low mass region}}, .

\bibitem{OurPaper4_2020}
S.~Banerjee, F.~Boudjema, N.~Chakrabarty and H.~Sun, \emph{{Relic density of
  dark matter in the inert doublet model beyond leading order: The SM Higgs
  resonance region}}, .

\bibitem{Boudjema:1995cb}
F.~Boudjema and E.~Chopin, \emph{{Double Higgs production at the linear
  colliders and the probing of the Higgs selfcoupling}},
  \href{http://dx.doi.org/10.1007/s002880050298}{\emph{Z. Phys. C} {\bf 73}
  (1996) 85--110}, [\href{http://arxiv.org/abs/hep-ph/9507396}{{\tt
  hep-ph/9507396}}].

\bibitem{Boudjema_2005}
F.~Boudjema, A.~Semenov and D.~Temes, \emph{Self-annihilation of the neutralino
  dark matter into two photons or azand a photon in the minimal supersymmetric
  standard model},
  \href{http://dx.doi.org/10.1103/physrevd.72.055024}{\emph{Physical Review D}
  {\bf 72} (Sep, 2005) }.

\bibitem{Baro:2007em}
N.~Baro, F.~Boudjema and A.~Semenov, \emph{{Full one-loop corrections to the
  relic density in the MSSM: A Few examples}},
  \href{http://dx.doi.org/10.1016/j.physletb.2008.01.031}{\emph{Phys. Lett.}
  {\bf B660} (2008) 550--560}, [\href{http://arxiv.org/abs/0710.1821}{{\tt
  0710.1821}}].

\bibitem{Baro:2008bg}
N.~Baro, F.~Boudjema and A.~Semenov, \emph{{Automatised full one-loop
  renormalisation of the MSSM. I. The Higgs sector, the issue of tan(beta) and
  gauge invariance}},
  \href{http://dx.doi.org/10.1103/PhysRevD.78.115003}{\emph{Phys. Rev.} {\bf
  D78} (2008) 115003}, [\href{http://arxiv.org/abs/0807.4668}{{\tt
  0807.4668}}].

\bibitem{Baro:2009na}
N.~Baro, F.~Boudjema, G.~Chalons and S.~Hao, \emph{{Relic density at one-loop
  with gauge boson pair production}},
  \href{http://dx.doi.org/10.1103/PhysRevD.81.015005}{\emph{Phys. Rev.} {\bf
  D81} (2010) 015005}, [\href{http://arxiv.org/abs/0910.3293}{{\tt
  0910.3293}}].

\bibitem{Boudjema:2011ig}
F.~Boudjema, G.~Drieu La~Rochelle and S.~Kulkarni, \emph{{One-loop corrections,
  uncertainties and approximations in neutralino annihilations: Examples}},
  \href{http://dx.doi.org/10.1103/PhysRevD.84.116001}{\emph{Phys. Rev. D} {\bf
  84} (2011) 116001}, [\href{http://arxiv.org/abs/1108.4291}{{\tt 1108.4291}}].

\bibitem{Boudjema:2014gza}
F.~Boudjema, G.~Drieu La~Rochelle and A.~Mariano, \emph{{Relic density
  calculations beyond tree-level, exact calculations versus effective
  couplings: the ZZ final state}},
  \href{http://dx.doi.org/10.1103/PhysRevD.89.115020}{\emph{Phys. Rev.} {\bf
  D89} (2014) 115020}, [\href{http://arxiv.org/abs/1403.7459}{{\tt
  1403.7459}}].

\bibitem{Belanger:2016tqb}
G.~B\'elanger, V.~Bizouard, F.~Boudjema and G.~Chalons, \emph{{One-loop
  renormalization of the NMSSM in SloopS: The neutralino-chargino and sfermion
  sectors}}, \href{http://dx.doi.org/10.1103/PhysRevD.93.115031}{\emph{Phys.
  Rev. D} {\bf 93} (2016) 115031}, [\href{http://arxiv.org/abs/1602.05495}{{\tt
  1602.05495}}].

\bibitem{Belanger:2017rgu}
G.~B\'elanger, V.~Bizouard, F.~Boudjema and G.~Chalons, \emph{{One-loop
  renormalization of the NMSSM in SloopS. II. The Higgs sector}},
  \href{http://dx.doi.org/10.1103/PhysRevD.96.015040}{\emph{Phys. Rev.} {\bf
  D96} (2017) 015040}, [\href{http://arxiv.org/abs/1705.02209}{{\tt
  1705.02209}}].

\bibitem{Banerjee:2019luv}
S.~Banerjee, F.~Boudjema, N.~Chakrabarty, G.~Chalons and H.~Sun, \emph{{Relic
  density of dark matter in the inert doublet model beyond leading order: The
  heavy mass case}},
  \href{http://dx.doi.org/10.1103/PhysRevD.100.095024}{\emph{Phys. Rev.} {\bf
  D100} (2019) 095024}, [\href{http://arxiv.org/abs/1906.11269}{{\tt
  1906.11269}}].

\bibitem{Bardin:1986fi}
D.~Bardin, S.~Riemann and T.~Riemann, \emph{{Electroweak One Loop Corrections
  to the Decay of the Charged Vector Boson}},
  \href{http://dx.doi.org/10.1007/BF01441360}{\emph{Z. Phys. C} {\bf 32} (1986)
  121--125}.

\bibitem{Belanger:2001fz}
G.~Belanger, F.~Boudjema, A.~Pukhov and A.~Semenov, \emph{{MicrOMEGAs: A
  Program for calculating the relic density in the MSSM}},
  \href{http://dx.doi.org/10.1016/S0010-4655(02)00596-9}{\emph{Comput. Phys.
  Commun.} {\bf 149} (2002) 103--120},
  [\href{http://arxiv.org/abs/hep-ph/0112278}{{\tt hep-ph/0112278}}].

\bibitem{Belanger:2004yn}
G.~Belanger, F.~Boudjema, A.~Pukhov and A.~Semenov, \emph{{micrOMEGAs: Version
  1.3}}, \href{http://dx.doi.org/10.1016/j.cpc.2005.12.005}{\emph{Comput. Phys.
  Commun.} {\bf 174} (2006) 577--604},
  [\href{http://arxiv.org/abs/hep-ph/0405253}{{\tt hep-ph/0405253}}].

\bibitem{Belanger:2006is}
G.~Belanger, F.~Boudjema, A.~Pukhov and A.~Semenov, \emph{{MicrOMEGAs 2.0: A
  Program to calculate the relic density of dark matter in a generic model}},
  \href{http://dx.doi.org/10.1016/j.cpc.2006.11.008}{\emph{Comput. Phys.
  Commun.} {\bf 176} (2007) 367--382},
  [\href{http://arxiv.org/abs/hep-ph/0607059}{{\tt hep-ph/0607059}}].

\bibitem{Belanger:2013oya}
G.~Belanger, F.~Boudjema, A.~Pukhov and A.~Semenov, \emph{{micrOMEGAs$\_$3: A
  program for calculating dark matter observables}},
  \href{http://dx.doi.org/10.1016/j.cpc.2013.10.016}{\emph{Comput. Phys.
  Commun.} {\bf 185} (2014) 960--985},
  [\href{http://arxiv.org/abs/1305.0237}{{\tt 1305.0237}}].

\bibitem{Belanger:2018mqt}
G.~Belanger, F.~Boudjema, A.~Goudelis, A.~Pukhov and B.~Zaldivar,
  \emph{{micrOMEGAs5.0 : Freeze-in}},
  \href{http://dx.doi.org/10.1016/j.cpc.2018.04.027}{\emph{Comput. Phys.
  Commun.} {\bf 231} (2018) 173--186},
  [\href{http://arxiv.org/abs/1801.03509}{{\tt 1801.03509}}].

\end{thebibliography}\endgroup
\bibliographystyle{JHEP}

\end{document}